\documentclass[journal=jacsat,manuscript=article]{achemso}  
\usepackage{amsfonts}
\usepackage{amssymb}
\usepackage{graphicx}
\usepackage{amsmath}
\usepackage{tikz}
\usepackage{adjustbox}
\usepackage[normalem]{ulem}
\usetikzlibrary{shapes.geometric, arrows}

\usepackage[english]{babel}
\usepackage{color}
\usepackage[version=3]{mhchem} 
\usepackage{natbib}
\usepackage{url}
\usepackage[colorlinks,citecolor=black,urlcolor=blue,linkcolor=black]{hyperref} 
\usetikzlibrary{arrows,shapes,positioning,shadows,trees}

\newcommand{\olr}[1]{{\color{red}{}}}

\newcommand{\osvp}[1]{{\color{red}{}}}

\author{Sebastian V. Pios}
\affiliation{Zhejiang Laboratory, Hangzhou 311100, China}
\author{Maxim F. Gelin}
\affiliation{School of Science, Hangzhou Dianzi University, Hangzhou 310018, China}
\author{Wolfgang Domcke}
\affiliation{School of Natural Sciences, Department of Chemistry, Technical University of Munich, D-85747 Garching, Germany}
\author{Lipeng Chen}
\affiliation{Zhejiang Laboratory, Hangzhou 311100, China}
\email{chenlp@zhejianglab.org}

\title{\textit{Ab Initio} Simulation of Femtosecond Time-Resolved Multi-Pulse Spectroscopies applied to the Heptazine$\cdots$H\textsubscript{2}O Complex
}

\begin{document}

\begin{abstract}
In multi-dimensional time-resolved spectroscopic experiments, multiple (more than two) short laser pulses with variable pulse delay times are employed for the time-resolved exploration of the photoinduced dynamics of molecular chromophores. 
In the present work, the quasi-classical doorway-window (DW) methodology recently developed for transient absorption pump-probe (PP) spectroscopy [M. F. Gelin et al., J. Chem. Theory Comput. 2021, 17, 2394] has been generalized to multi-pulse spectroscopies. 
Pump-push-probe (PPP) spectroscopy (involving three laser pulses) and pump-induced two-dimensional (P-2D) spectroscopy (involving five laser pulses) are considered as specific examples. 
The quasi-classical DW approximation results in conceptually simple and computationally efficient simulation protocols which are suitable for implementation with \textit{ab initio} on-the-fly electronic-structure calculations. 
Simulations of PPP and P-2D spectra performed for the hydrogen-bonded heptazine$\cdots$H\textsubscript{2}O complex illustrate that pump-stimulated experiments provide much richer information on the ultrafast radiationless relaxation dynamics of the excited electronic states of the heptazine$\cdots$H\textsubscript{2}O complex than conventional PP and 2D experiments.
\end{abstract}

Graphititic carbon nitride (\textit{g}-C\textsubscript{3}N\textsubscript{4})\cite{wang2009metal,wang_angewandte_2012},  as well as various related polymeric carbon nitrides,\cite{lau2022tour,kumar2023multifunctional,pelicano2024metal} have attracted much attention from the renewable energy research community as versatile photocatalysts for water splitting and for the oxidation of organic pollutants.\cite{ong_chemrev_2016,WEN2017} \textit{g}-C\textsubscript{3}N\textsubscript{4} is assumed to consist of heptazine (heptaazaphenalene) molecular building blocks which are connected by imine bonds.\cite{wang_angewandte_2012}
The poorly defined atomic structures and stoichiometries of these polymeric materials have been detrimental to the precise characterization of the primary reaction processes. The reaction mechanisms involved in the \textit{g}-C\textsubscript{3}N\textsubscript{4}-catalyzed evolution of hydrogen and oxygen from water or sacrificial reagents therefore are still a matter of debate.\cite{wang2009metal,domcke_chemphotochem2019,cruz2025carbon}

In an alternative approach, Schlenker and coworkers investigated the time-resolved spectroscopy and solution-phase photochemistry of a derivative of the heptazine (Hz) molecule. While the bare Hz molecule hydrolyzes rapidly in the presence of traces of water,\cite{halpern_hz_1984} the derivative trianisoleheptazine (TAHz) was found to be chemically and photochemically stable.\cite{rabe_tahz_2018} 
Under UV (365~nm) irradiation of TAHz in aqueous solution, the liberation of OH radicals was detected.\cite{rabe_tahz_2018} 
This finding supports the computational prediction of an excited-state proton-coupled electron-transfer (PCET) reaction in hydrogen-bonded TAHz$\cdots$H\textsubscript{2}O complexes, yielding TAHzH and OH radicals.\cite{ehrmaier_jpca_2020_molecular}
To obtain deeper insight into the mechanisms and ultrafast time scales of the photoinduced PCET in hydrogen-bonded complexes of Hz with protic substrates, \textit{ab initio} nonadiabatic classical trajectory simulations were performed by Huang and Domcke.\cite{xiang_heptazine2021} 
It was shown that excitation of the lowest bright $^1\pi\pi^*$ state of the Hz chromophore (the S\textsubscript{1}($\pi\pi^*$) state of Hz is a dark state) results in a rapid ($<$100~fs) radiationless relaxation through several dark $^1n\pi^*$ states to the S\textsubscript{1}($\pi\pi^*$) state, which is long-lived on this time scale. 
Experimental data for TAHz in liquid water indicate an S\textsubscript{1} lifetime of about 40~ns.\cite{rabe_tahz_2018} 
In competition with the ultrafast intramolecular relaxation within Hz, a charge-transfer (CT) state can promote the transfer of an electron and a proton from H\textsubscript{2}O to Hz along the hydrogen bond, resulting in HzH and OH radicals.\cite{ehrmaier_hz_2017} 
The branching ratio for H-atom transfer within 100~fs was estimated by these simulations to be about 10\%, while about 90\% of the electronic population were found to remain in the dark S\textsubscript{1} state.\cite{xiang_heptazine2021}

Using recently developed methodology for the computation of femtosecond time-resolved nonlinear electronic spectra from \textit{ab initio} on-the-fly nonadiabatic classical trajectories in the framework of the quasi-classical doorway-window (DW) approximation,\cite{OTFDW1,Xiang2D,gelin_WIRE2025,Carlos26} transient absorption (TA) pump-probe (PP) and two-dimensional (2D) electronic spectra were computed for the Hz$\cdots$H\textsubscript{2}O complex.\cite{pios_hz_h2o_pumpprobe}
These nonlinear signals consist of three contributions, ground-state bleach (GSB), stimulated emission (SE) and excited-state absorption (ESA). 
The strong SE signal from the bright $^1\pi\pi^*$ state reveals an ultrafast ($\approx$20~fs) radiationless decay of the $^1\pi\pi^*$ population as well as the continuous build-up of population in the S\textsubscript{1} state, beginning at about 40~fs (the time window of the simulation was 100~fs). 
The population of the nominally dark S\textsubscript{1} state becomes spectroscopically visible through vibronic intensity borrowing from the bright $^1\pi\pi^*$ state. 
The short-lived population of the bright $^1\pi\pi^*$ state also shows up as an intense signal in the ESA contribution to the TA PP signal. 
The most notable phenomenon in the 2D signals is an unusually large width of the ESA signal along the so-called emission frequency of the 2D spectrum, which reflects very large transient vibrational excess energy in the S\textsubscript{1} state immediately after relaxation from the bright $^1\pi\pi^*$ state.\cite{pios_hz_h2o_pumpprobe}
By comparison of the spectra of the Hz$\cdots$H\textsubscript{2}O complex with those of the bare Hz molecule, weak effects of the hydrogen bond on the ultrafast internal-conversion dynamics could be identified, albeit these spectra are primarily sensitive to the electronic relaxation dynamics within Hz and less so to the proton-transfer dynamics.\cite{pios_hz_h2o_pumpprobe}

The electron and proton dynamics induced by a time-delayed push pulse in the Hz$\cdots$H\textsubscript{2}O complex was studied with \textit{ab initio} trajectory simulations in another recent publication.\cite{pios_pumppush_md2025}
The push pulse was delayed by \textit{T}\textsubscript{1} = 100~fs relative to the pump pulse and the push-induced dynamics was simulated for another 100~fs. 
The re-excitation of the electronic population in the S\textsubscript{1} state by the push pulse populated higher excited states which are strongly nonadiabatically mixed with reactive CT states, resulting in significantly enhanced proton-transfer reactivity.\cite{pios_pumppush_md2025}

In the present work, the \textit{ab initio} simulation of pump-push-induced spectra for the Hz$\cdots$H\textsubscript{2}O complex has been completed by the inclusion of the respective probe pulses in the simulations. 
Various forms of multi-beam time-resolved spectroscopy were implemented in the past in different configurations.\cite{OBERLE1995,tahara2003,vanGrondelle2006,marek2011,kraack_pccp2013,buckup_annrev2014, kukura_physrev2020}. 
\begin{figure}
    \centering
    \includegraphics[width=0.5\linewidth]{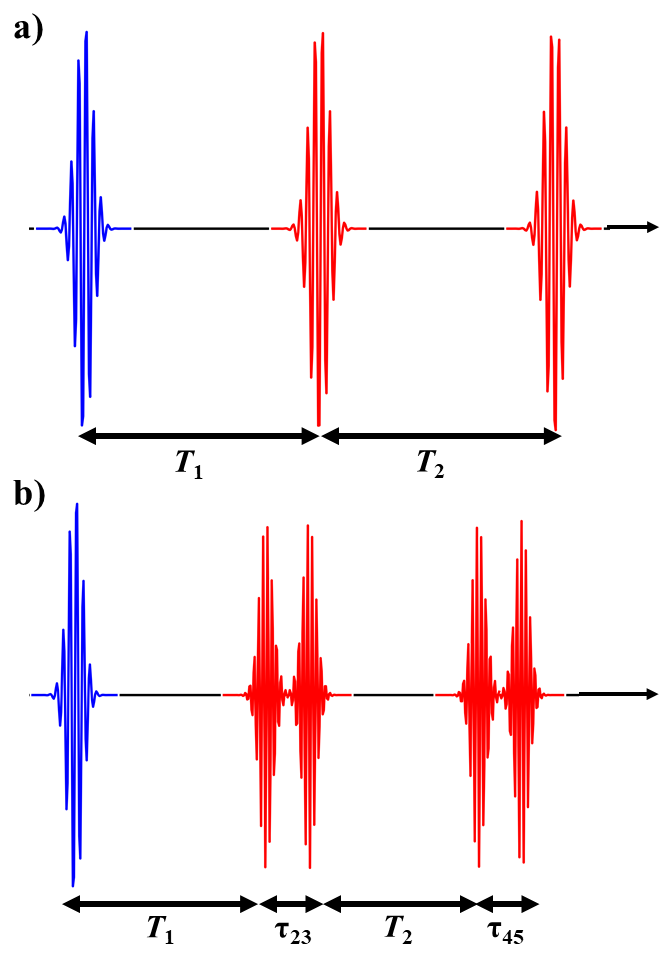}
    \caption{Scheme of pulse sequences and their respective time delays for three-beam pump-push-probe (a) and five-beam pump-2D (b) configurations. The pump pulse is shown in blue and the push and probe pulses are shown in red.}
    \label{fig:pulse_seq}
\end{figure}
In these spectroscopic techniques, an initial “actinic” pump pulse excites the sample from the electronic ground state to a dipole-allowed excited electronic state. In three-beam pump-push-probe (PPP) configurations (see Fig.~\ref{fig:pulse_seq}a), the push pulse at delay time \textit{T}\textsubscript{1} is followed by a probe pulse at delay time \textit{T}\textsubscript{2} which leads to self-heterodyne detection of the signal, analogous to TA PP spectroscopy. 
The two time delays \textit{T}\textsubscript{1} and \textit{T}\textsubscript{2} and the carrier frequencies of the pulses 2 and 3 can independently be chosen or varied, which results in numerous different realizations of PPP spectroscopy.\cite{buckup_annrev2014}
In addition, integral or dispersed detection of the probe signal\cite{MukamelBook} is possible. 
A different kind of multi-beam spectroscopy is fifth-order three-dimensional (3D) spectroscopy pioneered by Tan and collaborators.\cite{tan_jpcb2013,zhang2015direct}
In this technique, five laser pulses with three coherence intervals interact with the sample and generate a fifth-order signal.

In four- or five-beam experiments, a four-wave-mixing (FWM) measurement follows the actinic pump. 
These techniques are generally known as pump-FWM experiments.\cite{kraack_pccp2013,buckup_annrev2014} 
The most general experiment of this kind is the five-beam configuration illustrated in Fig.~\ref{fig:pulse_seq}b. 
The actinic pump is followed after a delay time \textit{T}\textsubscript{1} by a pair of phase-coherent excitation pulses separated by a time interval $\tau_{23}$. 
The free induction decay of the polarization generated by the pulses 2 and 3 is detected after a waiting time \textit{T}\textsubscript{2} by a second pair of phase coherent detection pulses separated by a time interval $\tau_{45}$. 
The last pulse usually is a so-called local oscillator which leads to heterodyne detection of the signal.\cite{kraack_pccp2013}
In this pump-2D (P-2D) experiment (simulated in the present work), the measured signal is Fourier transformed with respect to $\tau_{23}$ and $\tau_{45}$, which spreads the signal on two frequency axes.\cite{tian_science2003,jonas_annrev2003,cho2009two}

We present herein the first \textit{ab initio} simulations of PPP spectra and P-2D spectra for a molecular system. 
Among the many possible configurations mentioned above, we selected the integral PPP signal and the P-2D signal. 
Although the experimental realization of the integral PPP signal is more challenging than the implementation of the dispersed PPP signal, the integral detection provides time and frequency resolved signals over a much broader range of probe frequencies than the dispersed signal.\cite{MukamelBook} 
The frequency range of the latter is limited by the duration of the probe pulse. 
The pump-push time delay \textit{T}\textsubscript{1} is kept fixed at 100~fs, while the push-probe time delay \textit{T}\textsubscript{2} is varied. 
The five-beam P-2D signal provides the most complete information on the pump-induced dynamics by resolving the signal both with respect to the time-separation of the coherent excitation pulses and the time separation of the coherent detection pulses.
In this case, the time delay \textit{T}\textsubscript{1} also was kept fixed, while the time delay \textit{T}\textsubscript{2} was varied. 

PPP and P-FWM spectroscopic techniques were extensively applied by van Grondelle and coworkers and Buckup, Motzkus and coworkers for the exploration of the ultrafast photophysical dynamics of carotenoids, such as $\beta$-carotene.\cite{LARSEN2003,wohlleben2004,vanGrondelle2006,marek2011,buckup_annrev2014}
It is interesting to note that the photophysics of carotenoids and of Hz are indeed rather similar. 
In both chromophores, a very short-lived bright excited singlet state is populated by the pump pulse. 
In both cases, an ultrafast decay through conical intersections takes place and populates a long-lived S\textsubscript{1} state. The dark $^1n\pi^*$ states dispersed in between the $^1\pi\pi^*$ states of Hz turn out to be spectroscopically invisible. 
It therefore should make sense to apply the advanced spectroscopic techniques, which greatly enhanced our understanding of the photophysics of carotenoids, also to Hz and Hz-water complexes.
%
%

The photoinduced dynamics of the Hz$\cdots$H\textsubscript{2}O complex has been simulated with quasi-classical surface-hopping (SH) calculations.\cite{xiang_heptazine2021, pios_hz_h2o_pumpprobe} 
The energy of the electronic ground state has been calculated with the second-order M\o ller-Plesset (MP2) method.\cite{mp2_paper} 
The energies of the excited electronic states have been calculated with the algebraic-diagrammatic construction of second order (ADC(2)).\cite{schirmer1982} 
The electronic-structure calculations were performed with the TURBOMOLE package.\cite{turbomole} 
The correlation-consistent polarized split-valence double-$\zeta$ basis set (cc-pVDZ) was augmented with functions of the aug-cc-pVDZ basis\cite{dunning_basisset} on the atoms of the H\textsubscript{2}O molecule and the relevant N-atoms of Hz.\cite{pios_hz_h2o_pumpprobe}

The initial nuclear positions \textbf{R} and momenta \textbf{P} for the quasi-classical trajectory calculations were sampled from the zero-temperature harmonic Wigner distribution\cite{Wigner84} constructed with the MP2 Hessian. 
Following standard procedures,\cite{OTF2} the initial energies of the trajectories prepared by the pump pulse were stochastically selected from an energy window of width $\Delta$ = 0.1~eV centered at the maximum of the computed absorption spectrum ($\hbar\omega_{pu} = 4.29$~eV). 
This rectangular window  in the energy domain implies a pump pulse of the form $E_{pu}(t) =\sin(\Delta t/2)/(t/2)$ in the time domain with a duration of about 44~fs. 
N\textsubscript{1} = 300 trajectories generated by the pump pulse were propagated in manifold \{I\} of low-lying excited states (see Ref. \cite{OTFDW1} for the definition of the manifolds) with the velocity Verlet algorithm with a time step of 0.5~fs. 
M\textsubscript{I} = 8 excited electronic states were included in manifold \{I\}. 
Nonadiabatic transitions within manifold \{I\} were taken into account with a Landau-Zener (LZ) surface-hopping (LZSH) algorithm,\cite{LZM2,xie_pyrazine} employing the Belyaev-Lebedev formula for the hopping probabilities.\cite{belyaev2011} 
The trajectory simulations were performed with the ZagHop code.\cite{zaghop_code} 

For PPP spectroscopy, the push pulse was modeled by a Gaussian field envelope in the frequency domain with carrier frequency $\omega_{push}$ and width $\sigma_{push}$
\begin{equation}\label{eq:gauss_envelope}
    E_{push}(\omega) = \exp(-(\omega - \omega_{push})^{2} / \sigma_{push}^{2}).  
\end{equation}
The push pulse excites the electronic states \textit{e} of the trajectory ensemble at \textit{t} = \textit{T}\textsubscript{1} to manifold \{II\} of higher excited states \textit{f} with vertical excitation energies \textit{E\textsubscript{f}}(\textbf{R}) and transition dipole moments (TDMs) $\mu_{fe}$(\textbf{R}). 
M\textsubscript{II} = 22 higher excited states were included in manifold \{II\}. 
The initial conditions \textbf{R}, \textbf{P} of the trajectories prepared by the push pulse were sampled from the \textbf{R}, \textbf{P} of the pump-generated trajectory ensemble at \textit{t} = \textit{T}\textsubscript{1} and from the product of the power spectrum of the push pulse, $E^2_{push}(\omega)$, with the distribution of squared TDMs, $\mu_{fe}$(R)$^2$.
The carrier frequency ($\hbar\omega_{push}$ = 2.8~eV) and width ($\sigma_{push}$ = 3.7~eV), which corresponds to a duration of 1~fs of the push pulse, were chosen to optimize the overlap of $E_{push}(\omega)$ with the energy profile of the ESA spectrum calculated in Ref. \cite{pios_hz_h2o_pumpprobe}. 
N\textsubscript{2} = 287 trajectories prepared by the push pulse were propagated in the combined manifolds \{I\} and \{II\} from \textit{t} = \textit{T}\textsubscript{1}  = 100~fs up to \textit{t} = \textit{T}\textsubscript{2} = 200~fs.

The field envelope of the probe pulse in PPP spectroscopy also is of Gaussian form with carrier frequency $\omega_{pr}$ and width $\sigma_{pr}$

\begin{equation}\label{eq:gauss_envelope_probe}
    E_{pr}(\omega) = \exp(-(\omega - \omega_{pr})^{2} / \sigma_{pr}^{2}).  
\end{equation}
The carrier frequency and the width of the probe pulse are chosen as $\hbar\omega_{pr}$ = 2.8~eV and $\sigma_{pr}$ = 3.7~eV. 

In the quasi-classical DW approximation, TA PP signals are given as averages of the product of a doorway distribution (determined by the pump process) and three window functions $\mathcal{W}$\textsubscript{k}, k = GSB, SE, ESA (determined by the probe processes).
For the standard TA PP signal, the expressions for the doorway distribution and the functions $\mathcal{W}$\textsubscript{k} have been presented in the original publications\cite{OTFDW1,Xiang2D} and in a recent review\cite{gelin_WIRE2025}, to which the reader is referred. 
The corresponding equations for the novel multi-pulse signals evaluated in the present work are given in Section S1 of the SI.

For the calculation of the P-2D signals, the excitation of the pump-excited trajectory ensemble by the phase-coherent pulse pair 2, 3 at centered at \textit{t} = \textit{T}\textsubscript{1} is described by an analytically calculated doorway distribution which is analogous to the doorway distribution of standard 2D spectroscopy.\cite{GELIN2008,Xiang2D}
The effect of nuclear motion during the comparatively short time interval between pulses 2 and 3 is approximated by a phenomenological optical dephasing parameter $\nu$. 
The value of $\nu$ is chosen as 0.1~eV, which is a typical value for polyatomic molecules. 
The carrier frequencies of the excitation pulses ($\hbar\omega_2 = \hbar\omega_3$ = 2.8~eV) are the same as for the push pulse of PPP spectroscopy. 
The pulse duration is chosen as 0.1~fs.
This very short duration of the excitation and detection pulses is chosen to ensure a large window in the frequency domain. 
When longer pulses were chosen, the span of the spectra in the frequency domain would be correspondingly narrower. 
The initial conditions \textbf{R}, \textbf{P} of the trajectories prepared by the coherent excitation pulses were sampled from the positions and momenta of the pump-excited trajectory ensemble at \textit{t} = \textit{T}\textsubscript{1} and from the product of the power spectrum $E^2(\omega)$ of the excitation pulses and the distribution of squared TDMs.
N\textsubscript{2} = 287 trajectories prepared by the pair of excitation pulses were propagated in the combined manifolds \{I\} and \{II\} from \textit{t} = \textit{T}\textsubscript{1} = 100~fs up to \textit{t} = \textit{T}\textsubscript{2} = 200~fs. 
The window functions $\mathcal{W}_k$ are analogous to the window functions of standard 2D spectroscopy\cite{Xiang2D,gelin_WIRE2025} and are specified in Section 1 of the SI.

%
%
%
Turning to the results of the computational studies, the integral PPP spectrum of the Hz$\cdots$H\textsubscript{2}O complex consists of the SE and ESA contributions.
The GSB signal is negligible, because the
push pulse centered at $\hbar\omega_{push} = 2.8$~eV is out of resonance with the transitions to the ground state from the electronic levels around  $\hbar\omega_{pu} = 4.29$~eV excited by the pump pulse.  
In PPP spectroscopy, the delay time \textit{T}\textsubscript{2} of the probe pulse relative to the push pulse is analogous to the delay time \textit{T} in PP spectroscopy.  

\begin{figure}[h]
    \centering
    \includegraphics[width=1.0\linewidth]{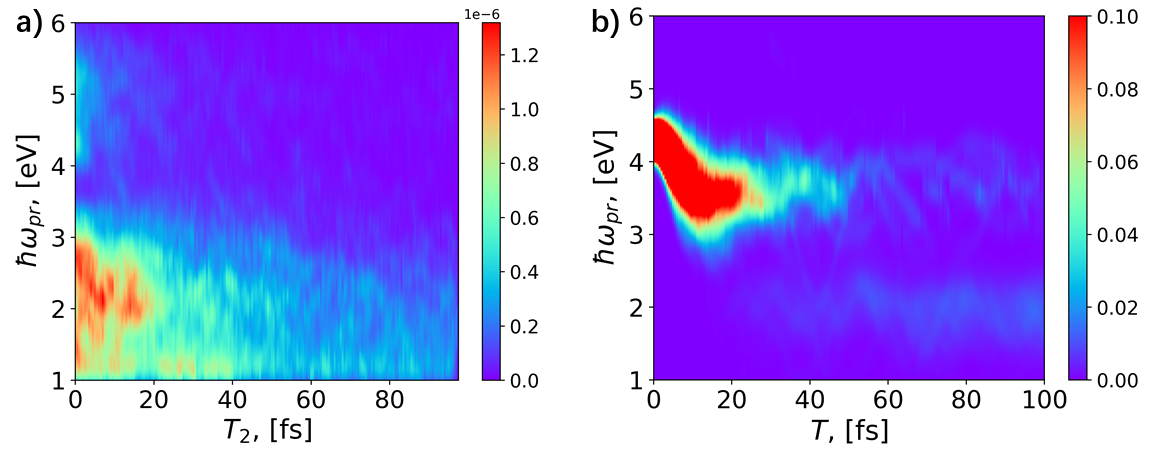}
    \caption{(a) SE contribution to the integral PPP signal of Hz$\cdots$H\textsubscript{2}O as a function of the push-probe delay time \textit{T}\textsubscript{2} and the probe carrier frequency $\hbar\omega_{pr}$. (b) SE contribution to the integral PP signal of Hz$\cdots$H\textsubscript{2}O as a function of the pump-probe delay time \textit{T} and the probe carrier frequency $\hbar\omega_{pr}$. The pulse durations of pump, push and probe pulses are 5~fs, 1~fs and 5~fs, respectively. The pump pulse is in resonance with the maximum of the UV absorption spectrum ($\hbar\omega_{pu}$ = 4.29~eV). The push pulse is in resonance with the maximum of the excited-state absorption spectrum ($\hbar\omega_{push}$ = 2.8~eV). The signal intensities are given relative to the maximum of the SE component of the TA PP signal. The data for (b) are taken from Ref.~\cite{pios_hz_h2o_pumpprobe}.}
    \label{fig:ta_se}
\end{figure}

The computed SE component of the integral PPP signal is displayed in Fig.~\ref{fig:ta_se}a as a function of \textit{T}\textsubscript{2} and $\hbar\omega_{pr}$.
At \textit{T}\textsubscript{2} = 0, an intense signal (seen in red in Fig.~\ref{fig:ta_se}a) is observed near $\hbar\omega_{pr}$ = 2.8~eV, coincident with the carrier frequency of the push pulse. 
This signal exhibits intensity oscillations in the delay time and shifts to the red in $\hbar\omega_{pr}$ before fading away at about 20~fs. 
A broader and weaker signal (seen in green and light blue in Fig.~\ref{fig:ta_se}a) also exhibits a red shift and fades away on a time scale of about 60~fs. 
At very short delay times \textit{T}\textsubscript{2}, one can recognize SE signals near 4.2~eV and 5.0~eV. 
These high-frequency signals fade away within just a few femtoseconds.

To put this PPP SE signal into perspective, we compare it with the integral PP SE signal of Hz$\cdots$H\textsubscript{2}O, taken from Ref. 20 and displayed in Fig.~\ref{fig:ta_se}b. 
The PP SE signal images the rapid ($\approx$ 20~fs) radiationless decay of the population of the bright $^1\pi\pi^*$ state of Hz (seen in red) through invisible $^1n\pi^*$ states to the S\textsubscript{1}($\pi\pi^*$) state. 
The population of the latter is visible at $\hbar\omega_{pr}\approx$ 2.0~eV in Fig.~\ref{fig:ta_se}b due to intensity borrowing from the bright $^1\pi\pi^*$ state. 

The very short-lived signal at $\hbar\omega_{pr}\approx$ 4-5~eV in Fig.~\ref{fig:ta_se}a arises from SE from the higher excited states prepared by the push pulse down to the electronic ground state. 
This weak signal reflects the extremely short lifetime of these states and their small electronic transition moments and Franck-Condon factors to the electronic ground state (several of these states are dark CT states from H\textsubscript{2}O to Hz).

The most intense signal in Fig.~\ref{fig:ta_se}a, located at $\hbar\omega_{pr}\approx$ 2.8~eV and \textit{T}\textsubscript{2} $\approx$ 0, arises from stimulated emission from the vibrationally hot population of the low-lying $^1\pi\pi^*$ states generated by the ultrafast decay of the higher excited states \textit{f} prepared by the push pulse. 
The lifetime and the red shift of this signal are qualitatively similar to lifetime and red-shift of the SE signal of the bright $^1\pi\pi^*$ state in Fig.~\ref{fig:ta_se}b. 
It should be kept in mind that the condition of validity of the DW approximation (non-overlapping pulses) is not fulfilled at pulse delay times of a few femtoseconds. 
The signals therefore cannot be expected to agree quantitatively with experimental signals for these very short pulse delay times. 
The latter typically exhibit a coherent artifact for overlapping pump and probe pulses.\cite{kraack_pccp2013}

The comparatively long-lived background signal seen in green and blue in Fig.~\ref{fig:ta_se}a, which gradually recedes to lower $\hbar\omega_{pr}$, reflects stimulated emission from intermediate electronic states in the relaxation cascade which have low TDMs with the lower electronic states. 
Overall, the signal in Fig.~\ref{fig:ta_se}a reflects the initially extremely fast and subsequently slower radiationless return of the electronic population prepared by the push pulse to the long-lived S\textsubscript{1} state of the Hz$\cdots$H\textsubscript{2}O complex.

In the calculations, it is possible to extract the SE signal just from the S\textsubscript{1} state of Hz. 
This signal is shown in Fig.~S1a in the SI. 
It may be compared with the SE signal from the S\textsubscript{1} state in the PP spectrum (Fig.~\ref{fig:ta_se}b). 
The comparison reveals that the re-population of the S\textsubscript{1} state via radiationless relaxation from the higher electronic states prepared by the push pulse is delayed in comparison with the radiationless relaxation from the bright $^1\pi\pi^*$ state in PP spectroscopy. 
The PCET reactivity (H-atom transfer from H\textsubscript{2}O to Hz), which is enhanced by the push pulse,\cite{pios_pumppush_md2025} is not directly visible in the PP and PPP signals which mainly reveal the dynamics of electronic state populations of Hz.

\begin{figure}[h]
    \centering
    \includegraphics[width=1.0\linewidth]{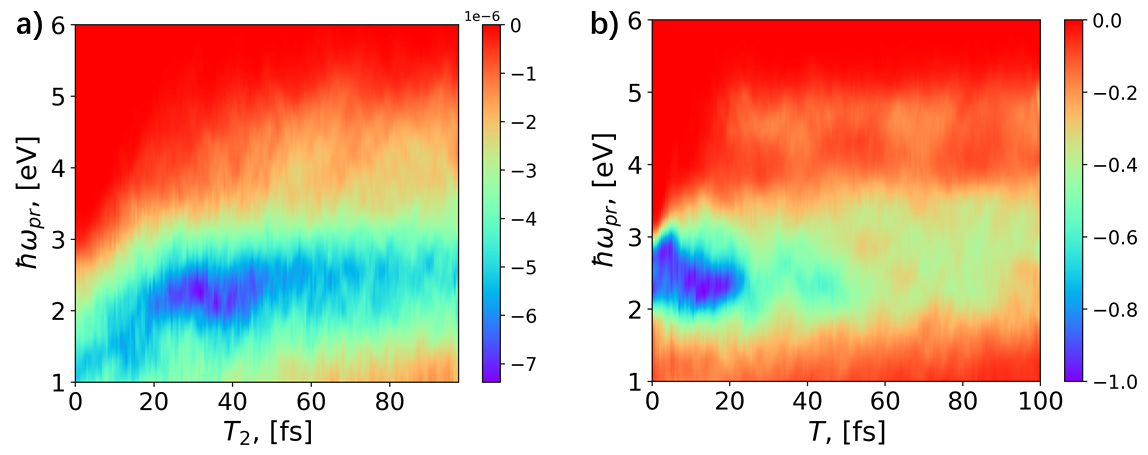}
    \caption{(a) ESA contribution to the integral PPP signal of Hz$\cdots$H\textsubscript{2}O as a function of the push-probe delay time \textit{T}\textsubscript{2} and the probe carrier frequency $\hbar\omega_{pr}$. (b) ESA contribution to the integral PP signal of Hz$\cdots$H\textsubscript{2}O as a function of the pump-probe delay time \textit{T} and the probe carrier frequency $\hbar\omega_{pr}$. The pulse durations of pump, push and probe pulses are 5~fs, 1~fs and 5~fs, respectively. The pump pulse is in resonance with the maximum of the UV absorption spectrum ($\hbar\omega_{pu}$ = 4.29~eV). The push pulse is in resonance with the maximum of the excited-state absorption spectrum ($\hbar\omega_{push}$ = 2.8~eV). The (negative) signal intensities are given relative to the maximum of the ESA  component of the TA PP signal. The data for (b) are taken from Ref.~\cite{pios_hz_h2o_pumpprobe}.}
    \label{fig:ta_esa}
\end{figure}

The computed ESA contribution to the TA PPP signal of Hz$\cdots$H\textsubscript{2}O is shown in Fig.~\ref{fig:ta_esa}a. 
At \textit{T}\textsubscript{2} = 0~fs, a broad signal is seen which extends from $\hbar\omega_{pr}$ = 2.0~eV to frequencies below 1.0~eV. 
This signal shifts to the blue within 20~fs and its intensity increases, culminating at \textit{T}\textsubscript{2} $\approx$ 30~fs. 
At longer delay times \textit{T}\textsubscript{2}, the signal becomes quasi-stationary with moderate intensity fluctuations. 
For comparison, the integral TA PP signal of Hz$\cdots$H\textsubscript{2}O, taken from Ref. 20, is shown in Fig.~\ref{fig:ta_esa}b. 

The intense PP ESA signal located between 0 and 20~fs and 2.2 and 2.9~eV in Fig.~\ref{fig:ta_esa}b has been assigned to absorption from the degenerate S\textsubscript{5}/S\textsubscript{6} state of Hz (the bright $^1\pi\pi^*$ state) to higher excited states.\cite{pios_hz_h2o_pumpprobe}
While the intensity of the PP ESA signal fades away after about 30~fs (Fig.~\ref{fig:ta_esa}b), the intensity of the PPP ESA signal remains approximately constant (Fig.~\ref{fig:ta_esa}a). 
A closer analysis of the contributions of trajectories reveals that the bright $^1\pi\pi^*$ state becomes continuously repopulated by radiationless relaxation from higher excited states. 
In addition to the extremely fast initial decay of the highest excited states prepared by the push pulse, there exists a cascade of slower excited-state relaxation which extends beyond 100~fs.

In the trajectory simulations, the PPP ESA signal arising just from the long-lived S\textsubscript{1} state of Hz can be separated from the total PPP ESA signal. 
This signal is displayed in Fig.~S1b in the SI. 
It reveals that the re-population of the S\textsubscript{1} state starts at about 40~fs and increases up to 100~fs. 
The signal becomes very broad in $\hbar\omega_{pr}$ with increasing \textit{T}\textsubscript{2}, reflecting the formation of a hot pre-equilibrium distribution of vibrational levels in the long-lived S\textsubscript{1} state.
%

%
Turning to the {-2D signal, the SE component of the P-2D signal of the Hz$\cdots$H\textsubscript{2}O complex with re-phasing phase-matching condition is displayed in Fig.~\ref{fig:2d_se_all}a-d as function of excitation and detection frequencies for four selected waiting times \textit{T}\textsubscript{2} = 20, 40, 60, 80~fs. 
The signals exhibit unexpectedly rich peak structures. 
The intense peaks (with positive signal) are all located below the diagonal, that is, the emission frequencies of the peaks are lower than the excitation frequencies. 
The intense peaks are found in an about 1~eV-wide band centered at the carrier frequency of the excitation pulse pair (2.8~eV). 
The extension of the peak structure in the emission frequency is wider, about 3.0~eV. 
Part of the signal is cut off by the lower limit of the emission frequency in the panels of Fig.~\ref{fig:2d_se_all}a-d. 
Weaker and less structured negative signals are seen in the upper left part and at the bottom of the panels around an excitation frequency of $\approx$ 4~eV. 
With increasing waiting time \textit{T}\textsubscript{2}, the intensity located at the excitation frequency of $\approx$ 2.8~eV moves to lower emission frequencies, which is reminiscent of the red shift of the SE component of the PPP signal (Fig.~\ref{fig:ta_se}a). 
Clearly, the P-2D SE signal contains much richer dynamical information than the PPP SE signal.

It is instructive to compare the P-2D SE signals with the corresponding 2D signals of the Hz$\cdots$H\textsubscript{2}O complex calculated in Ref. 20. 
These signals are displayed in Fig.~\ref{fig:2d_se_all}e-h for the same waiting times. 
At short waiting times, the 2D SE signal is narrow in both excitation and emission frequencies. 
While the 2D signal remains narrow in the excitation frequency with increasing waiting time, it broadens significantly in the emission frequency (Fig.~\ref{fig:2d_se_all}e-h). 
The narrow distribution in the excitation frequency is the consequence of the compact distribution of \textbf{R} and \textbf{P} in the electronic ground state. 
The remarkably wide distribution of the P-2D SE signals, on the contrary, reflects the fact that the excitation at \textit{T}\textsubscript{1} = 100~fs occurs from a broad distribution of trajectories over \textbf{R}, \textbf{P} as well as over several excited electronic states. 
The trajectory ensemble which has been formed by radiationless relaxation from the bright $^1\pi\pi^*$ state of Hz is thus directly imaged by the P-2D SE signal. 
It is noteworthy that the broad distribution of intensities is already fully developed at \textit{T}\textsubscript{2} = 20~fs, illustrating the ultrafast decay of the bright $^1\pi\pi^*$ state to lower-lying states in a very short time. Fig.~\ref{fig:2d_se_all}a illustrates ultrafast radiationless decay through conical intersections caught in action.

\begin{figure}[H]
    \centering
    \includegraphics[width=0.5\linewidth]{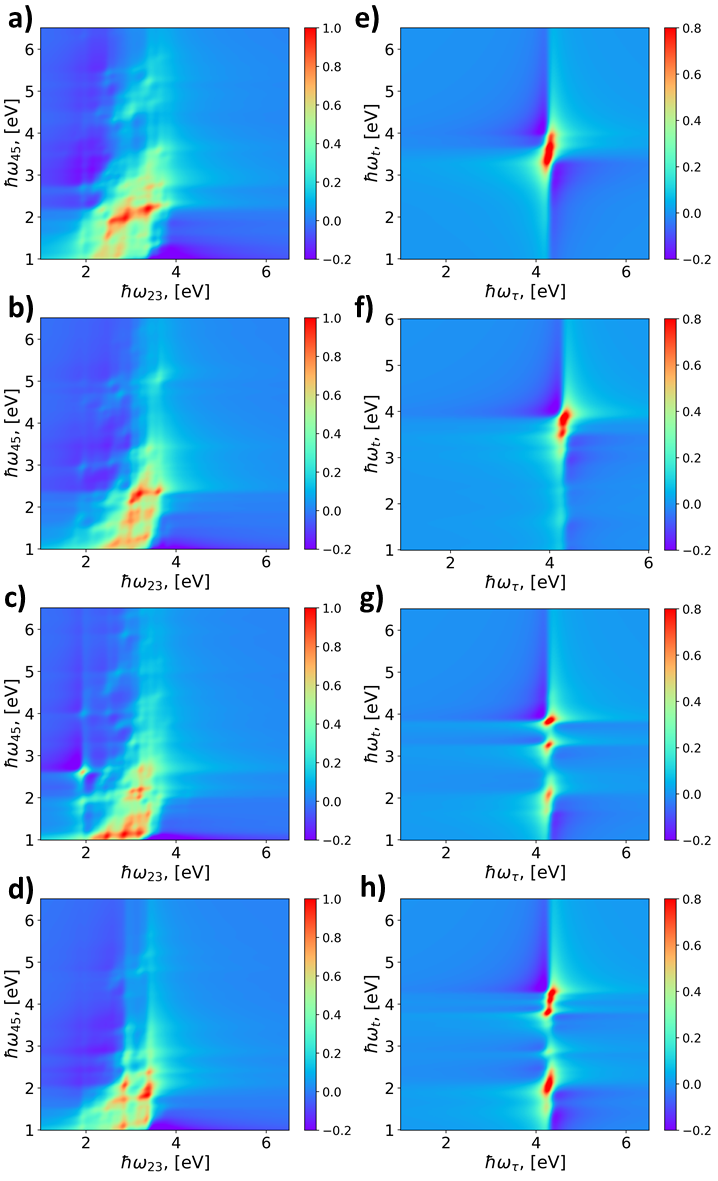}
    \caption{(a-d) SE contribution to the rephasing P-2D signal of Hz$\cdots$H\textsubscript{2}O as a function of the excitation and detection frequencies at waiting times \textit{T}\textsubscript{2} = 20, 40, 60, 80~fs. (e-h) SE contribution to the rephasing 2D signal of Hz$\cdots$H\textsubscript{2}O as a function of excitation and detection frequencies at waiting times \textit{T} = 20, 40, 60, 80~fs. The pulse durations of excitation and detection pulses are 0.1~fs. The signals for each waiting time have been rescaled to the same maximum intensity for better visibility.}
    \label{fig:2d_se_all}
\end{figure}

The ESA component of the P-2D signal of the Hz$\cdots$H\textsubscript{2}O complex is presented in Fig.~\ref{fig:2d_esa_all}a-d for the same four waiting times, \textit{T}\textsubscript{2} = 20, 40, 60, 80~fs. 
At \textit{T}\textsubscript{2} = 20~fs, the peak of the (negative) signal is centered near 2.8~eV in the excitation frequency and near 2.4~eV in the emission frequency. 
While the peak location is stationary in the excitation frequency, the location decreases toward 2.2~eV in the emission frequency for longer waiting times (Fig.~\ref{fig:2d_esa_all}a-d). 
It should be noted that the orientation of the elongated peak evolves from near-horizontal to vertical from \textit{T}\textsubscript{2} = 20~fs to \textit{T}\textsubscript{2} = 80~fs (Fig.~\ref{fig:2d_esa_all}a-d). 
The width of the main signal is about 1.5~eV in the excitation frequency and about 4~eV in the emission frequency (ignoring the tail of the emission below 1.0~eV). 

For comparison, the corresponding 2D ESA signals of the Hz$\cdots$H\textsubscript{2}O complex are presented in Fig.~\ref{fig:2d_esa_all}e-h. 
Analogous to the SE signal, the P-2D ESA signals are much wider in the excitation frequency than the 2D ESA signals, reflecting the fact that the ESA transitions in the P-2D signal occur from a broad distribution of electronic states and nuclear geometries of the pump-excited ensemble. 
Due to the high density of upper excited states, the P-2D ESA signal is less structured than the P-2D SE signal.

The relatively constant intensity of the P-2D ESA signal as function of the waiting time \textit{T}\textsubscript{2} results from a more or less constant population of the electronic states S\textsubscript{4}, S\textsubscript{5}, S\textsubscript{6} of the Hz chromophore which are particularly bright in absorption. 
While the populations of these states relax quickly to lower electronic states, they are continuously regenerated by the decay of higher electronic states which were populated by the excitation pulse pair, resulting in an transiently quasi-stationary population of the S\textsubscript{4}, S\textsubscript{5}, S\textsubscript{6} states. 

\begin{figure}[H]
    \centering
    \includegraphics[width=0.5\linewidth]{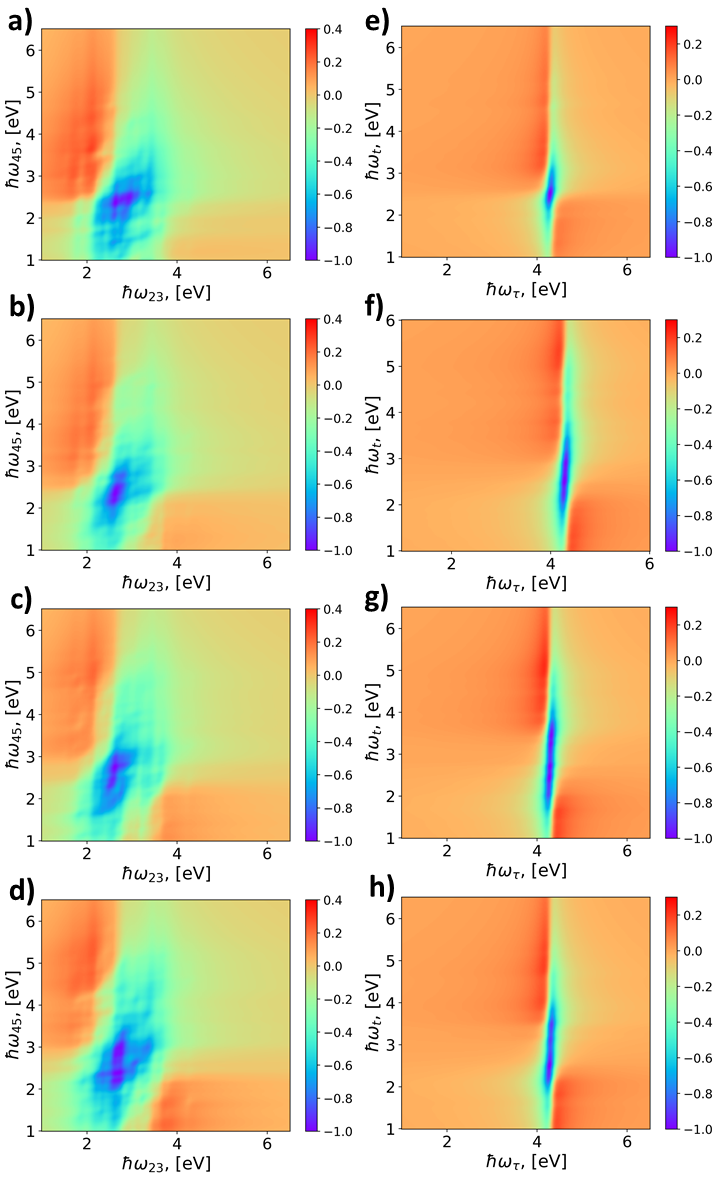}
    \caption{(a-d) ESA contribution to the rephasing P-2D signal of Hz$\cdots$H\textsubscript{2}O as a function of the excitation and detection frequencies at waiting times \textit{T}\textsubscript{2} = 20, 40, 60, 80~fs. (e-h) ESA contribution to the rephasing 2D signal of Hz$\cdots$H\textsubscript{2}O as a function of excitation and detection frequencies at waiting times \textit{T} = 20, 40, 60, 80~fs. The pulse durations of excitation and detection pulses are 0.1~fs. The signals for each waiting time have been rescaled to the same maximum intensity for better visibility.}
    \label{fig:2d_esa_all}
\end{figure}

In summary, we performed \textit{ab initio} on-the-fly trajectory simulations of PPP and pump-2D experiments in the framework of the quasi-classical DW approximation for the example of the Hz$\cdots$H\textsubscript{2}O complex. 
In the Hz chromophore, the population of the bright $^1\pi\pi^*$ state by the actinic pump pulse triggers radiationless relaxation within tens of femtoseconds, resulting in a vibrationally hot quasi-stationary ensemble in the long-lived S\textsubscript{1} state. 
In the PPP experiment, the push pulse re-excites the hot ensemble in the S\textsubscript{1} state to higher electronic states and the ensuing chain of radiationless transitions is recorded by the probe pulse. 
In the P-2D experiment, a FWM measurement is performed on the hot ensemble in the S\textsubscript{1} state with pairs of phase-coherent excitation and detection pulses. 
The conventional Franck-Condon principle as well as the electronic symmetry selection rules applying for excitation from the electronic ground state are largely relaxed in the hot S\textsubscript{1} ensemble. 
Therefore, the PPP and P-2D spectra contain much more detailed information on the radiationless relaxation dynamics of the short-lived $^1\pi\pi^*$ state of Hz as well as higher excited states than the corresponding experiments performed on the Hz chromophore in the electronic ground state.

Whereas the calculation of transient signals via the evaluation of the fifth-order polarization or higher-order polarizations is very tedious and may hardly be possible for complex molecular systems, the quasi-classical DW approximation provides a conceptually intuitive and computationally efficient protocol for the evaluation of multi-pulse signals within its range of validity (short and non-overlapping laser pulses). 
The accuracy of the sampling of the nonadiabatic dynamics by classical trajectories and additional simplifications made in the DW approximation may require further exploration and improvements in the future. 
The most crucial issue, however, likely is the accuracy of the \textit{ab initio} electronic-structure calculations, especially for the higher excited electronic states.

A challenge in time-resolved multi-pulse spectroscopic experiments is the multitude of possibilities to choose the pulse parameters and to vary the pulse delay times. 
First-principles simulations of femtosecond time-resolved nonlinear signals for various experimental configurations could assist in the planning and optimization of sophisticated spectroscopic experiments in the future. 
Moreover, the simulations yield the GSB, SE and ESA components of the signals separately, while experimentally only the sum of the three components is accessible. 
By disentangling the three contributions, first-principles simulations may be of great help in the analysis and interpretation of the measured signals.

\section{Supporting Information}
Equations for PP and PPP signals, as well as for 2D and P-2D signals. Spectroscopic signals arising exclusively from the S\textsubscript{1} state
\begin{acknowledgement}
S.V.P. acknowledges support from the National Natural Science Foundation of China (No.~W2433024). 
L.P.C. acknowledges support from the National Natural Science Foundation of China (No.~ 22473101).
M.F.G. acknowledges support from the National Natural Science Foundation of China (No.~22373028). 
\end{acknowledgement}
\section{Data availability statement}
The data that supports the findings of this study is available from the corresponding author upon reasonable request.
%
%
%
%
\section{Competing interests}
The authors declare that they have no competing interests.
%
%
\newpage
\bibliography{main}
\end{document}